\documentclass[twocolumn,showpacs,preprintnumbers,amsmath,amssymb]{revtex4}
\usepackage{graphicx}
\usepackage{dcolumn}
\usepackage{bm}
\begin{document} 
\title{Quasiparticle Band Structure and Spin Excitation Spectrum of the 
Kondo Lattice}
\author{R. Eder}
\affiliation{Karlsruhe Institute of Technology,
Institut f\"ur Festk\"orperphysik, 76021 Karlsruhe, Germany}
\date{\today}
\begin{abstract}
A formulation of the Kondo lattice Hamiltonian in terms of bond particles
is derived and solved in two different approximations. The bond particles 
correspond to the eigenstates of a single unit cell
and are bosons for states with even electron number and fermions
for states with an odd electron number. As a check various 
physical quantities
are calculated for the 1D Kondo insulator and good agreement with
numerical results is obtained for $J/t>1$.
\end{abstract} 
\pacs{71.27.+a,75.30.Mb,71.28.+d}
\maketitle
\section{Introduction}
Metallic compounds containing Cerium, Ytterbium or Uranium - the so-called 
Heavy  fermions - continue to be a much studied field of solid state physics. 
These materials show a number of remarkable phenomena which are widely believed
to be caused by the strong Coulomb repulsion between the electrons in the 
$4f$-shells of Cerium and Ytterbium or the $5f$-shell of Uranium. A long known 
phenomenon is the crossover 
from a lattice of localized $f$ electrons coexisting with weakly or moderately
correlated conduction electron bands at high temperature, to an exotic Fermi 
liquid with strongly correlation-enhanced effective masses at low temperature, 
whereby the $f$ electrons now contribute to the Fermi-surface 
volume\cite{Stewart,kondoinsulators}.
The low temperature Fermi-liquid phase can undergo magnetic ordering 
transitions whereby the transition temperature often can be 
tuned to zero by external parameters
such as temperature, pressure, magnetic field or alloying, resulting in 
quantum critical points and the ensuing non-Fermi-liquid behavior
and superconducting domes\cite{StewartII,loenireview,Steglichreview}. 
An intriguing feature thereby is the fact that whereas the $f$ electrons do 
contribute to the Fermi surface volume in the paramagnetic phase they seem
to `drop out' of the Fermi-surface volume at some of these transitions.\\
Heavy fermion compounds can be described by the
Kondo-lattice model (KLM). In the simplest case of no orbital degeneracy
each unit cell $n$ contains one $f$ orbital and one conduction band orbital,
and denoting the creation operators for electrons in these orbitals by
$f_{n,\sigma}^\dagger$ and $c_{n,\sigma}^\dagger$ the Kondo lattice Hamiltonian 
is $H=H_t + H_J$ with
\begin{eqnarray}
H_t&=&\sum_{m,n}\;\sum_\sigma\; t_{m,n}\;c_{m,\sigma}^\dagger c_{n,\sigma}^{},
\nonumber \\
H_J&=&\sum_n\;h_n,\nonumber \\
h_n&=& J\;\bm{S}_{n,f}\cdot \bm{S}_{n,c},\nonumber\\
\bm{S}_{n,c} &=&\frac{1}{2}\;\sum_{\sigma,\sigma'}\;  c_{n,\sigma}^\dagger\; 
\bm{\tau}_{\sigma,\sigma'}\;
 c_{n,\sigma'}^{}.
\label{h_i}
\end{eqnarray}
Here $\bm{\tau}$ denotes the vector of Pauli matrices 
(an analogous definition holds for $\bm{S}_{n,f}$).
An important feature of the KLM is the constraint to have precisely one 
electron per $f$ orbital:
\begin{eqnarray}
\sum_\sigma\;f_{n,\sigma}^\dagger f_{n,\sigma}^{} &=& 1,
\label{fconst}
\end{eqnarray}
which must hold separately for each $n$. In the following
we consider a lattice with $N$ unit cells and $N_c$ conduction electrons.
The KLM can be derived from the more realistic periodic
Anderson model (PAM) by means of the Schrieffer-Wolff 
transformation\cite{SchriefferWolff}.\\
The impurity versions of the Kondo and Anderson model are well 
understood. Approximate solutions can be obtained by
variational wave functions\cite{Yosida,Varma,Gu},
mean-field (or saddle point) approximation to the
exchange term\cite{YoshimoriSakurai,ReadNewns},
or Green's function techniques where the hybridization or exchange between 
the $f$ level and conduction band are treated as 
perturbation\cite{Keiter,Kuramoto,Coleman}.
Thereby the inverse of the degeneracy $n_f$ of the $f$ level, $1/n_f$, 
often plays the role of a small parameter\cite{Bickers}. Exact solutions of the
impurity models can be obtained by renormalization group\cite{RG} 
and Bethe ansatz\cite{Bethe}. 
The ground state of the impurity  is a singlet formed from
the $f$ electron on the impurity and an extended screening state
formed from the conduction band states, whereby for weak coupling ($J\ll t$) 
the binding energy - the so-called Kondo temperature - is 
$k_BT_K\propto We^{-\frac{1}{\rho J}}$. Here $W$ and $\rho$ are
bandwidth and density of states of the conduction band.\\
The lattice versions of the model are less-well understood.
A noteworthy result is the fact that even for the KLM,
where the $f$ electrons are strictly localized, they do contribute
to the Fermi surface volume\cite{Oshikawa} provided the
system is a Fermi liquid. Approximate results for the KLM 
and PAM have been obtained in the mean-field (or saddle-point)
approximation. For the KLM the exchange term, which is quartic in electron 
operators is 
factorized\cite{LacroixCyrot,Lacroix,AuerbachLevin,Burdinetal,ZhangYu,Lavagna,Senthil,Global,ZhangSuLu,Nilsson,Montiel}, whereas for the PAM a 
slave-boson representation is 
used\cite{MillisLee,ReadNewns1}. The models also have been studied
using Gutzwiller-type trial wave functions\cite{RiceUeda,Fazekas}.
The resulting band structure is consistent
with a simple hybridization picture: a dispersionless effective $f$ band
close to the Fermi energy of the decoupled conduction band
hybridizes with the conduction electron band via an effective
hybridization matrix element $\propto k_B T_K$ at weak coupling.
This results in a Fermi surface with a volume corresponding to
itinerant $f$ electrons and the `heavy bands' characteristic of
heavy fermion compounds. 
In addition to mean-field theories, a large amount of quantitative
results has also been gathered by numerical methods such as 
Density Matrix Renormalization Group (DMRG)\cite{yuwhite,MC1,MC2,Mutou,Smerat},
Dynamical Mean-Field (DMFT) calculations\cite{J1,J2,J3,peterspruschke},
Quantum Monte-Carlo (QMC)\cite{Carsten,Assaad}, 
series expansion (SE)\cite{series,seriesexp,Trebst},
Variational Monte-Carlo (VMC)\cite{WatanabeOgata,Asadzadeh,Kubo},
exact diagonalization\cite{Tsunetsugu,White,Tsutsui},
Dynamical Cluster Approximation\cite{MartinAssaad,MartinBerxAssaad}
and Variational Cluster Approximation\cite{Lenz}.
For the paramagnetic phase the numerical techniques produce band 
structures which are consistent with the hybridization picture, 
whereby it has to be kept in mind that numerical methods often have
problems to access the limit of small $J/t$ and thus to reproduce the
Kondo scale $k_BT_K$. However, the heavy quasiparticles and the fact that
the $f$ electrons do participate to the Fermi surface in the KLM and PAM
are reproduced.\\
Considerable effort was devoted to a study of the magnetic phase 
transitions which are believed to be due to a competition\cite{Doniach} 
between the singlet formation in the impurity model and the
RKKY-interaction between the $f$ spins\cite{RKKY} mediated by the conduction 
electrons. A controversial question is whether
the heavy quasiparticles persist at the magnetic transition, so that
this may be viewed as the heavy Fermi liquid undergoing a conventional
spin-density-wave transition, or whether the magnetic ordering
suppresses the heavy Fermi liquid alltogether. Numerous studies have addressed 
magnetic ordering\cite{K1,K2,K3,K4,K5,K6,K7,K8,K9,K10,K11,K12,K13}
but open questions remain.\\
It is the purpose of the present manuscript to present a theory for
the single-particle band structure and spin excitation spectrum of the
Kondo lattice which relies on the interpretation of the eigenstates of
a single cell as fermionic or bosonic particles, which we call bond particles.
Bond particle theory was proposed originally by 
Sachdev and Bhatt\cite{SachdevBhatt} to study
spin systems and applied to spin ladders\cite{Gopalan},
bilayers\cite{vojta1,vojta2}, intrisically dimerized spin 
systems\cite{sushkov,Park} and the `Kondo necklace'\cite{Siahatgar}.
It was also applied to the PAM\cite{Oana}, as well as 
antiferromagnetic (AF) ordering in the planar 
KLM\cite{JureckaBrenig,KotovHirschfeld}, 
a discussion of its different AF phases\cite{afbf} and the band structure
in the AF phase\cite{afbf1}.
It is by nature a strong-coupling theory which should work best in the
(unphysical) limit of $J/t\gg 1$. However, as will be shown below by comparison
to numerical results, there is some reason to hope that the theory retains its
validity down to $t/J\approx 1$ which may be sufficient to discuss
magnetic ordering phenomena.
\section{Hamiltonian}
We consider eigenstates of the single cell exchange term
$h_n$. Introducing the matrix $4$-vector
$\bm{\gamma}=(\tau_0,\bm{\tau})$ with $\tau_0=1$
the state $4$-vector $\bm{\beta}_n=(s_n, \bm{t}_n)$ 
is\cite{SachdevBhatt,Gopalan}
\begin{eqnarray}
|\bm{\beta}_n\rangle =\frac{1}{\sqrt{2}}\;\sum_{\sigma,\sigma'}\;
c_{n,\sigma}^\dagger \;(\bm{\gamma}i\tau_y)_{\sigma,\sigma'}\;f_{n,\sigma'}^\dagger
|0\rangle.
\label{bosons}
\end{eqnarray}
These are the
singlet ($|s_n\rangle$) with energy
$-\frac{3J}{4}$ and the three components of the triplet
($|\bm{t}_n\rangle$) with energy $\frac{J}{4}$.\\
The single-cell states with an odd number of electrons 
(which have energy $0$) are 
\begin{eqnarray}
|a,n,\sigma\rangle &=& f_{n,\sigma}^\dagger|0\rangle ,\nonumber \\
|b,n,\sigma\rangle &=& c_{n,\uparrow}^\dagger c_{n,\downarrow}^\dagger
f_{n,\sigma}^\dagger|0\rangle.
\label{fermions}
\end{eqnarray}
We now rewrite the KLM as a Hamiltonian for
bosons and fermions which correspond to these single-cell eigenstates.
More precisely, if a given cell $n$ is in one of the states
(\ref{bosons}) with $2$ electrons we consider it as occupied by a boson, 
created by the respective operator $4$-vector $(s_n^\dagger,\bm{t}_n^\dagger)$ 
whereas if it is in one of the states (\ref{fermions}) with a single (three) 
electrons we consider it as occupied by a fermion created by 
$a_{n,\sigma}^\dagger$ ($b_{n,\sigma}^\dagger$). The latter correspond
to the `bachelor spins' in the $U/t=\infty$ Hubbard model
to which the KLM reduces\cite{Nozieres} for $J/t\rightarrow \infty$  and 
$N_c \ne N$. \\
In order for this representation to make sense each cell must be occupied
by precisely one of these particles resulting in the constraint
(to be obeyed for each $n$)
\begin{eqnarray}
s_n^\dagger s_n^{} + \sum_\sigma \left(\;a_{n,\sigma}^\dagger a_{n,\sigma}^{} +
b_{n,\sigma}^\dagger b_{n,\sigma}^{} \right)+ \bm{t}_n^\dagger \cdot \bm{t}_n^{} = 1.
\label{constraint_1}
\end{eqnarray}
On the other hand, each of the basis states (\ref{bosons}) and
(\ref{fermions}) obeys the constraint (\ref{fconst}) exactly,
so that this constraint is `built in' into the theory.
In terms of the bond particles the exchange term is
\begin{eqnarray}
H_J&=&\sum_n\;\left(\; \frac{3J}{4}\sum_\sigma\;
(b_{n,\sigma}^\dagger b_{n,\sigma}^{} + a_{n,\sigma}^\dagger a_{n,\sigma}^{})
+ J \;\bm{t}_n^\dagger \cdot \bm{t}_n^{}\;\right)\nonumber \\
&&\;\;\;\;\;\;\;\;\;\;\;\;\;\;\;\;\;\;\;\;\;\;\;\;\;\;\;\;\;\;\;\;
 - \frac{3NJ}{4}.
\label{HJ}
\end{eqnarray}
On the other hand, we might also write
\begin{eqnarray}
H_J&=&\sum_n\;\left(\; -\frac{3J}{4}\;s_n^\dagger s_n^{}
+ \frac{J}{4} \;\bm{t}_n^\dagger \cdot \bm{t}_n^{}\;\right).
\label{HJ1}
\end{eqnarray}
As long as the constraint (\ref{constraint_1}) holds these two
forms are equivalent - we wil continue to use (\ref{HJ}).
From now on we take a fermion operator with omitted
spin index to denote a two-component column vector, e.g.
\begin{eqnarray*}
c_n&=&\left(\begin{array}{c}
c_{n,\uparrow} \\
c_{n,\downarrow}
\end{array} \right).
\end{eqnarray*}
In this notation the representation of the electron annihilation
operator in terms of the bond particles is
\begin{eqnarray}
c_n=\frac{1}{\sqrt{2}}\;:\left(\;(s_n+ {\bm t}_n\cdot{\bm \tau})\;
i\tau_y a_n^\dagger - (s_n^\dagger - {\bm t}_n^\dagger\cdot {\bm \tau})\;
 b_n\;\right):\nonumber \\
\label{cop}
\end{eqnarray}
where $:\dots:$ denotes normal ordering.
Up to the numerical prefactors the form of this equation
follows from the requirement that both sides be covariant
spinors and the fact that ${\bm t}$ and ${\bm t}^\dagger$ 
are vector operators. 
The representation of the kinetic energy $H_t$ is obtained by 
substituting (\ref{cop}) into (\ref{h_i}). One obtains
$H_t=H_1+H_2+H_3+H_4$ with
\begin{widetext}
\begin{eqnarray}
H_1&=&\sum_{m,n}\;\frac{t_{m,n}}{2}\;\sum_\sigma \;\left(\;b_{m,\sigma}^\dagger b_{n,\sigma}^{} -
a_{m,\sigma}^\dagger a_{n,\sigma}^{}\;\right)\;s_n^\dagger\; s_m^{} 
-\sum_{m,n}\;\frac{t_{m,n}}{2}\; 
\left[\;\left(\;
b_{m,\uparrow}^\dagger a_{n,\downarrow}^\dagger - b_{m,\downarrow}^\dagger a_{n,\uparrow}^\dagger
\;\right)\;s_m^{}\; s_n^{} +H.c.\;\right],\nonumber\\
H_2&=&\sum_{m,n}\;\frac{t_{m,n}}{2}\;\sum_\sigma \;\left(\;b_{m,\sigma}^\dagger b_{n,\sigma}^{} -
a_{m,\sigma}^\dagger a_{n,\sigma}^{}\;\right)\;\bm{t}_n^\dagger\cdot \bm{t}_m^{}\;
+\sum_{m,n}\;\frac{t_{m,n}}{2}\;
\left[\;\left(\;
b_{m,\uparrow}^\dagger a_{n,\downarrow}^\dagger
- b_{m,\downarrow}^\dagger a_{n,\uparrow}^\dagger
\;\right) \bm{t}_m^{}\cdot \bm{t}_n^{}\;+ H.c\;\right],\nonumber\\
H_3&=& -\;\sum_{m,n}\;\frac{t_{m,n}}{2}\;\left[\;\left(\;{\bm \pi}^\dagger_{m,n}\cdot
\left(\;s_m^{}\;\bm{t}_n^{} - \bm{t}_m^{}\;s_n^{}\right) + H.c.\;\right)
+\left(\;\bm{b}_{m,n} - \bm{a}_{m,n}\;\right)\cdot
(\bm{t}_n^\dagger \;s_m^{} + s_n^\dagger\;\bm{t}_m^{}\;)\;\right],\nonumber\\
H_4&=&
\sum_{m,n}\;\frac{t_{m,n}}{2}\;\left[\;\left[\;i{\bm \pi}^\dagger_{m,n}\cdot
(\;\bm{t}_m^{}\times\bm{t}_n^{}\;)
+ H. c. \;\right]
- i\;(\;\bm{b}_{m,n} - \bm{a}_{m,n}\;)\cdot
(\;\bm{t}_n^\dagger\times \bm{t}_m^{}\;)\;\right],
\label{hamtot}
\end{eqnarray}
\end{widetext}
with the following vectors formed from the fermions:
\begin{eqnarray*}
\bm{b}_{m,n}&=& \sum_{\sigma,\sigma'}\;b_{m,\sigma}^\dagger\;\bm{\tau}_{\sigma,\sigma'}\;b_{n,\sigma'}^{},
\nonumber \\
\bm{a}_{m,n}&=& \sum_{\sigma,\sigma'}\;a_{m,\sigma}^\dagger\;\bm{\tau}_{\sigma,\sigma'}\;a_{n,\sigma'}^{},
\nonumber \\
\bm{\pi}^\dagger_{m,n}&=& \sum_{\sigma,\sigma'}\;b_{m,\sigma}^\dagger\;
\left(\bm{\tau} \;i\tau_y\right)_{\sigma,\sigma'}\;a_{n,\sigma'}^\dagger.
\end{eqnarray*}
Strictly speaking the individual terms in this Hamiltonian
have to be `site-wise normal ordered' e.g.
$b_{m,\sigma}^\dagger b_{n,\sigma}^{} t_{n,\alpha}^\dagger t_{m,\alpha}^{}
\rightarrow b_{m,\sigma}^\dagger t_{m,\alpha}^{} t_{n,\alpha}^\dagger b_{n,\sigma}^{}$
but since this normal ordering always involves commutation of
a fermion and a boson neither nonvanishing commutators nor
Fermi signs will arise.
The number of electrons - {\em including the localized
$f$ electrons} - in the system is
\begin{eqnarray}
N_e&=&2\;\sum_n\;(\;s_n^\dagger s_n^{} + \bm{t}_n^\dagger \cdot \bm{t}_n^{}\;)
\nonumber \\
&&\;\;\;\;\;\;\;\;\;\;\;\;\;\;\;\;
+\sum_{n,\sigma} \left(\;3\;b_{n,\sigma}^\dagger b_{n,\sigma}^{} + a_{n,\sigma}^\dagger 
a_{n,\sigma}^{} \right)\nonumber \\
&=&2N + \sum_n\;\sum_\sigma \left(\;b_{n,\sigma}^\dagger b_{n,\sigma}^{} - 
a_{n,\sigma}^\dagger a_{n,\sigma}^{}
\right)\nonumber \\
&=&  \sum_n\;\sum_\sigma \left(\;b_{n,\sigma}^\dagger b_{n,\sigma}^{} + 
a_{n,\sigma}^{} a_{n,\sigma}^\dagger\;\right),
\label{constraint_2}
\end{eqnarray}
where (\ref{constraint_1}) was used to obtain the second line.
The Hamiltonian (\ref{HJ})$+$(\ref{hamtot}) together with the 
constraint (\ref{constraint_1}) provides an exact representation of the
KLM. On the other hand, it
is compliated and impossible to solve even approximately e.g. by diagrammatic
methods, due to the constraint (\ref{constraint_1}),
which is equivalent to an infinitely strong Hubbard-like
repulsion between the bond particles. The whole formulation
in terms of bond particles will be useful only if we can identify
the fermions and triplet bosons as approximate quasiparticles and
spin excitations of the system and find a way 
to extract a sufficiently simple yet accurate theory for these.
A considerable simplification becomes possible by making use of
the fact - to be verified below - that over large regions of parameter space 
the densities of fermions and triplets are relatively small so that the vast
majority of cells is in the singlet state. Thus, if one can get
rid of the singlets by either considering them as
condensed or by re-interpreting the singlet as the 
`true vacuum state' of a cell, one retains a theory for a system of
fermions and bosons which in principle are still subject to the infinitely 
strong repulsion implied by the constraint (\ref{constraint_1}) but which have
a low density so that relaxing the constraint may be a reasonable approximation.
Put another way, by using the bond particles one can trade
the constraint (\ref{fconst}) which refers to a dense system of electrons - the
density of $f$ electrons is 1/cell - for a constraint like
(\ref{constraint_1}) without singlets
which refers to a system of particles with a relatively low density.
In the following, we explore two possible approximation 
schemes to  `get rid of the singlets' and compare the results to numerical 
calculations. It should also be noted that while we will not
do so in the following, it is in principle
possible to deal with strong repulsion in a low density
system by well-known field theoretical methods\cite{FW}. For
the case of bond bosons in spin systems this has in fact been carried
out explicitely by Kotov {\em et al.}\cite{Kotov} and 
Shevchenko {\em et al.}\cite{Shevchenko}.
We will compare the results from bond particle theory to numerical results
for the paramagnetic state in a 
one-dimensional chain with only nearest neighbor hopping $-t$ and
$n_e=2$ - i.e. the 1D Kondo insulator\cite{uedareview},
throughout $t$ is the unit of energy.
\section{Mean Field Theory}
As a first approximation we study the Hamiltonian in mean-field
approximation. This approximation was applied previously to spin 
systems\cite{SachdevBhatt,Gopalan} and to antiferromagnetic ordering in the 
planar KLM\cite{JureckaBrenig}.
Since we are interested in the paramagnetic phase
we initially drop the terms $H_3$ and $H_4$.
$H_3$ describes  pair creation and propagation processes
whereby a single triplet-boson is absorbed or emitted. In mean-field
theory this term would contribute only in a state where the triplets
are condensed\cite{JureckaBrenig} i.e. a magnetically ordered 
state\cite{SachdevBhatt}.
Similarly, $H_4$ describes pair creation and propagation processes
whereby two triplets coupled to a vector are emitted/absorbed. The resulting 
vector-like order parameters would be important to describe a state with
incommensurate or spiral magnetic order but vanish in a rotationally invariant 
state.\\
In the remaining terms $H_J+H_1+H_2$ the singlets are assumed to be condensed 
whence the corresponding operators can be replaced by a real number,
$s_n^\dagger,s_n^{} \rightarrow s$. The condensation amplitude
$s$ now is a freely variable internal parameter of the system, to be 
determined by minimization of the Helmholtz free energy.
The constraint  (\ref{constraint_1}) is replaced by the global
constraint
\begin{eqnarray}
\sum_n\;\left(\;s^2 + \bm{t}_n^{\dagger}\cdot \bm{t}_n^{} +
\sum_\sigma (b_{n,\sigma}^\dagger b_{n,\sigma}^{} + a_{n,\sigma}^\dagger a_{n,\sigma}^{})
\;\right)&=&N.\nonumber \\
\label{constraint_g}
\end{eqnarray}
We perform a Hartree-Fock factorization of the
quartic terms in $H_2$ and add the constraints
(\ref{constraint_g}) and (\ref{constraint_2})
using the Lagrange multipliers $\lambda$ and $\mu$, respectively.
We call the resulting Hamiltonian $H_{MF}$ and have
$H_{MF} = H_F+ H_B + Nc$. The fermionic Hamiltonian is
\begin{eqnarray*}
H_F&=&  e_0\;\sum_{n,\sigma}\;\left(\;b_{n,\sigma}^\dagger b_{n,\sigma}^{} +
a_{n,\sigma}^\dagger a_{n,\sigma}^{}\;\right)\nonumber\\
&&\;\;+\sum_{m,n}\;\tilde{t}_{m,n}\;\sum_\sigma\;\left(\;b_{m,\sigma}^\dagger 
b_{n,\sigma}^{} -
a_{m,\sigma}^\dagger a_{n,\sigma}^{}\;\right)\nonumber \\
&&\;\;\;\;-\sum_{m,n}\;\left[\Delta_{m,n}\;\left(\;
b_{m,\uparrow}^\dagger a_{n,\downarrow}^\dagger
- b_{m,\downarrow}^\dagger a_{n,\uparrow}^\dagger
\;\right) + H.c.\right]\\
&&\;\;\;\;-\mu\; \sum_{n,\sigma}\;\left(\;b_{n,\sigma}^\dagger b_{n,\sigma}^{} +
a_{n,\sigma}^{} a_{n,\sigma}^\dagger\;\right)\\
\end{eqnarray*}
where $e_0=\frac{3J}{4}-\lambda$ and
\begin{eqnarray*}
\tilde{t}_{m,n}&=& \frac{t_{m,n}}{2} s^2 +\zeta_{m,n}\\
\Delta_{m,n}&=&\frac{t_{m,n}}{2} s^2 + \eta_{m,n}.
\end{eqnarray*}
with
 $\zeta_{m,n}=\frac{t_{m,n}}{2}\;\langle \bm{t}_n^\dagger\cdot 
\bm{t}_m^{}\;\rangle$,
$\eta_{m,n}=-\frac{t_{m,n}}{2}\;\langle \bm{t}_m^{}\cdot \bm{t}_n^{}\;\rangle$.
We consider a translationally invariant and isotropic state and accordingly
assume that expectation values such as $\zeta_{m,n}$ depend only on
$|\bm{R}_m - \bm{R}_n|$ so that
\begin{eqnarray*}
H_F &=& e_0\;\sum_{\bm{k},\sigma}\;\left(\;b_{\bm{k},\sigma}^\dagger 
b_{\bm{k},\sigma}^{} -a_{-\bm{k},\bar{\sigma}}^{} a_{-\bm{k},\bar{\sigma}}^\dagger\;\right)\\
&&+ \sum_{\bm{k},\sigma}\;(\tilde{t}_{\bm{k}} -\mu)
\left(\; b_{\bm{k},\sigma}^\dagger b_{\bm{k},\sigma}^{} +
a_{-\bm{k},\bar{\sigma}}^{} a_{-\bm{k},\bar{\sigma}}^\dagger
\;\right) \\
&&\;\;\;\;\;\;\;\;-\sum_{\bm{k},\sigma}\;sign(\sigma)(\Delta_{\bm{k}}\;
b_{\bm{k},\sigma}^\dagger 
a_{-\bm{k},\bar{\sigma}}^\dagger + H.c.)\\
&&\;\;\;\;\;\;\;\;\;\;\;\;\;\;\;\; -\;2\sum_{\bm{k}} \;(\tilde{t}_{\bm{k}} -e_0), 
\end{eqnarray*}
with the Fourier transform of the hopping integral
\begin{eqnarray}
\tilde{t}_{\bm{k}}&=&\sum_{\bm{r}}\;e^{i \bm{k}\cdot \bm{r}}\;
\tilde{t}_{\bm{r}}\nonumber \\
&=& \sum_{\bm \alpha}\;z_{\bm \alpha} \tilde{t}_{\bm \alpha} 
\gamma_{\bm \alpha}({\bm k})
\label{foudef}
\end{eqnarray}
and an analogous definition of $\Delta_{\bm{k}}$.
Here ${\bm \alpha}$ denotes shells of symmetry-equivalent
neighbors of a given site, $z_{\bm \alpha}$ the number of neighbors
belonging to a shell, and $\gamma_{\bm \alpha}$ the respective tight-binding
harmonic.
This can be solved by the unitary transformation
\begin{eqnarray}
\alpha_{\bm{k}}^\dagger &=&\;\;\,
u_{\bm{k}}\;b_{\bm{k}}^\dagger + v_{\bm{k}}\;
i\tau_y\;a_{-\bm{k}}^{},\nonumber \\
\beta_{\bm{k}}^\dagger &=&
-v_{\bm{k}}\;b_{\bm{k}}^\dagger + u_{\bm{k}}\;
i\tau_y\;a_{-\bm{k}}^{},
\label{unitary}
\end{eqnarray}
so that
\begin{eqnarray*}
H_F&=&\sum_{\bm{k},\sigma}\;(\;E_{\alpha,\bm{k}}\; \alpha_{\bm{k},\sigma}^\dagger 
\alpha_{\bm{k},\sigma}^{}
+ E_{\beta,\bm{k}}\; \beta_{\bm{k},\sigma}^\dagger \beta_{\bm{k},\sigma}^{}\;)\\
&& \;\;\;\;\;\;\;\;\;\;\;\;\;\;\;\;\;\;\;\;\;\;\;\;\;\;\;\;\;\;\;\; 
-\;2\sum_{\bm{k}} \;(\tilde{t}_{\bm{k}} -e_0).
\end{eqnarray*}
Here $E_{\nu,\bm{k}}=\tilde{t}_{\bm{k}} \pm W_{\bm{k}} - \mu$
and $\alpha$ corresponds to the lower of the two energies.
Thereby
\begin{eqnarray*}
W_{\bm{k}}&=&\sqrt{ e_0^2 + \Delta_{\bm{k}}^2},\\
u_{\bm{k}}&=&-\sqrt{\frac{W_{\bm{k}}-e_0}{2W_{\bm{k}}}},\\
v_{\bm{k}}&=&\frac{-\Delta_{\bm{k}}}{\sqrt{2W_{\bm{k}}(W_{\bm{k}}-e_0)}}.
\end{eqnarray*}
By virtue of the unitarity of (\ref{unitary}) it follows that the
electron number (\ref{constraint_2}) becomes
\begin{eqnarray*}
N_e&=&2\;\sum_{{\bm k},\sigma}\;(\;\alpha_{\bm{k},\sigma}^\dagger \alpha_{\bm{k},\sigma}^{}
+ \beta_{\bm{k},\sigma}^\dagger \beta_{\bm{k},\sigma}^{}\;).
\end{eqnarray*}
The volume of the quasiparticle Fermi surface therefore corresponds
to both, conduction electrons and $f$ electrons.
Despite the fact that all basis states have precisely
one $f$ electron per cell, so that these are strictly localized, 
the $f$ electrons do
contribute to the Fermi surface volume as if they  were 
itinerant\cite{Oshikawa}.\\
The bosonic Hamiltonian is (with $\tilde{J}=J-\lambda$)
\begin{eqnarray*}
H_B&=&\tilde{J}\;\sum_n \;\bm{t}_n^\dagger \cdot \bm{t}_n^{}
+ \sum_{m,n}\;\tilde{\zeta}_{m,n} \;\bm{t}_n^\dagger \cdot \bm{t}_m^{}\\
&&\;\;\;\;\;\;\;\; + \sum_{m,n}(\;\tilde{\eta}_{m,n}\;\bm{t}_m^{} \cdot \bm{t}_n^{} + H.c.),\\
\tilde{\zeta}_{m,n} &=& \frac{t_{m,n}}{2}\;\sum_\sigma \;
\langle \;b_{m,\sigma}^\dagger b_{n,\sigma}^{} - a_{m,\sigma}^\dagger a_{n,\sigma}^{}\;\rangle,\\
\tilde{\eta}_{m,n} &=&  \frac{t_{m,n}}{2}\;\langle \;b_{m,\uparrow}^\dagger a_{n,\downarrow}^\dagger
- b_{m,\downarrow}^\dagger a_{n,\uparrow}^\dagger\; \rangle.
\end{eqnarray*}
Fourier transformation gives
\begin{eqnarray*}
H_B&=&\sum_{\bm{k}}\;(\tilde{J}+\tilde{\zeta}_{\bm{k}})
\;\bm{t}_{\bm{k}}^\dagger \cdot 
\bm{t}_{\bm{k}}^{}
 + \sum_{\bm{k}}\;( \tilde{\eta}_{\bm{k}}\;\bm{t}_{\bm{k}}^{} \cdot \bm{t}_{-\bm{k}}^{}
+ H.c.),
\end{eqnarray*}
with $\tilde{\zeta}_{\bm{k}}$ and $\tilde{\eta}_{\bm{k}}$ defined
as in (\ref{foudef}).
This can be solved by the ansatz
$\bm{\tau}_{\bm{k}}^\dagger= \tilde{u}_{\bm{k}} \;\bm{t}_{\bm{k}}^\dagger + \tilde{v}_{\bm{k}} \;\bm{t}_{-\bm{k}}^{}$
and $H_B$ becomes
\begin{eqnarray*}
H_B&=&\sum_{\bm{k}}\;\omega_{\bm{k}} \;\bm{\tau}_{\bm{k}}^\dagger \cdot 
\bm{\tau}_{\bm{k}}^{}
+ \frac{3}{2}\;\sum_{\bm{k}}\;(\omega_{\bm{k}} - (\tilde{J} + \tilde{\zeta}_{\bm{k}})).
\end{eqnarray*}
Thereby
\begin{eqnarray*}
\omega_{\bm{k}}&=&\sqrt{(\tilde{J}+\tilde{\zeta}_{\bm{k}})^2 - 4\;\tilde{\eta}_{\bm{k}}^2}\\
\tilde{u}_{\bm{k}}&=&\frac{2\tilde{\eta}_{\bm{k}}}
{\sqrt{{2\omega_{\bm{k}}(\tilde{J}+\tilde{\zeta}_{\bm{k}}-\omega_{\bm{k}})}}},\\
\tilde{v}_{\bm{k}}&=&
\sqrt{\frac{\tilde{J}+\tilde{\zeta}_{\bm{k}}-\omega_{\bm{k}}}{{2\omega_{\bm{k}}}}}.
\end{eqnarray*}
The additive constant is $Nc$ with
\begin{eqnarray*}
c&=&-\frac{3J}{4}+\lambda (1-s^2)+\mu n_e -\Theta,\nonumber \\
\Theta&=&\sum_{\bm{\alpha}}\;\frac{2z_{\bm{\alpha}}}{t_{\bm{\alpha}}}\;
\left(\; \zeta_{\bm{\alpha}} \tilde{\zeta}_{\bm{\alpha}} - 2
 \eta_{\bm{\alpha}} \tilde{\eta}_{\bm{\alpha}}\;\right),
\end{eqnarray*}
and the Helmholtz free energy becomes
\begin{widetext}
\begin{eqnarray*}
F &=& -\frac{2}{\beta}\;\sum_{\bm{k}}\;\sum_{\nu\in\{\alpha,\beta\}}\;
\log\left(1+e^{-\beta E_{\nu,\bm{k}}}\right)
+ \frac{3}{\beta}\;\sum_{\bm{k}}\;\log\left(1-e^{-\beta \omega_{\bm{k}}}\right) 
 -\;2\sum_{\bm{k}} (\tilde{t}_{\bm{k}} - e_0) 
+ \frac{3}{2}\;\sum_{\bm{k}}\;
\left(\omega_{\bm{k}} - (\tilde{J}+\tilde{\zeta}_{\bm{k}})\right)
+ Nc.
\end{eqnarray*}
Had we used the alternative form (\ref{HJ1}) for $H_J$ we would have obtained
the same expression with $\lambda-\frac{3J}{4} \rightarrow \lambda$.
Minimizing $F$ with respect to $\lambda$ gives
$1 - s^2 - n_F - n_B =0$ where
\begin{eqnarray*}
n_F &=&\frac{1}{N}\sum_{\bm{k},\sigma}\; \langle b_{\bm{k},\sigma}^\dagger 
b_{\bm{k},\sigma}^{} + a_{\bm{k},\sigma}^\dagger a_{\bm{k},\sigma}^{}\rangle 
=\frac{2}{N}\sum_{\bm{k}}\left[ 1 - \frac{e_0}{W_{\bm{k}}}\;
\left(\;f(E_{\alpha,\bm{k}}) - f(E_{\beta,\bm{k}})\;\right)\right],\nonumber \\
n_B &=& \frac{1}{N}\sum_{\bm{k}}\;\langle 
\bm{t}^\dagger \cdot \bm{t}^{} \rangle 
= \frac{3}{N}\sum_{\bm{k}}\;\left[\frac{\tilde{J} + \tilde{\zeta}_{\bm{k}}}{2\omega_{\bm{k}}}\;\coth\left(\frac{\beta \omega_{\bm{k}}}{2} \right)
-\frac{1}{2}\right],
\end{eqnarray*}
\end{widetext}
are the densities of fermions and bosons, respectively.
Minimization with respect to the $\zeta$ and $\eta$ parameters
in $H_F$ and $H_B$ gives the self-consistency equations
\begin{eqnarray*}
\zeta_{\bm{\alpha}}&=& 
\frac{3t_{\bm{\alpha}}}{2N}\sum_{\bm{k}}\;\gamma_{\bm{\alpha},\bm{k}}
\left[\frac{\tilde{J}+\tilde{\zeta}_{\bm{k}}}{2\omega_{\bm{k}}}
\coth\left(\frac{\beta \omega_{\bm{k}}}{2}\right) - \frac{1}{2}\right],\\
\eta_{\bm{\alpha}}&=& 
\frac{3t_{\bm{\alpha}}}{2N}\sum_{\bm{k}}\;\gamma_{\bm{\alpha},\bm{k}} \;
\frac{\tilde{\eta}_{\bm{k}}}{\omega_{\bm{k}}}
\coth\left(\frac{\beta \omega_{\bm{k}}}{2}\right),\\
\tilde{\zeta}_{\bm{\alpha}}&=& \frac{t_{\bm{\alpha}}}{N}
\;\sum_{\bm{k}}\;\gamma_{\bm{\alpha},\bm{k}} \;\left[
f(E_{\alpha,\bm{k}}) + f(E_{\beta,\bm{k}}) -1\right],\\
\tilde{\eta}_{\bm{\alpha}}&=& \frac{t_{\bm{\alpha}}}{N}
\;\sum_{\bm{k}}\;\gamma_{\bm{\alpha},\bm{k}} \;
\frac{\Delta_{\bm{k}}}{2W_{\bm{k}}}\;
\left[\;f(E_{\alpha,\bm{k}}) - f(E_{\beta,\bm{k}})\;\right].
\end{eqnarray*}
The above equations can also
be derived `directly', by evaluating the respective
thermal averages with the bosonic and fermionic mean-field Hamiltonian.
Differentiation with respect to $s$ gives the additional condition
\begin{eqnarray*}
\lambda &=&
\frac{1}{N}\;
\sum_{\bm{k}}\;\epsilon_{\bm{k}} \;\left[\;f(E_{\alpha,\bm{k}}) + 
f(E_{\beta,\bm{k}}) -1\;\right]
\nonumber \\
&& \;\;\;\;\;\;\;\;- \frac{1}{N}\;
\sum_{\bm{k}}\;\epsilon_{\bm{k}}\;\frac{\eta_{\bm{k}}}{W_{\bm{k}}}\;
\left[\;f(E_{\alpha,\bm{k}}) - f(E_{\beta,\bm{k}})\;\right],
\end{eqnarray*}
where $\epsilon_{\bf k}$ is the noninteracting dispersion.
The resulting set of coupled equations can be solved numerically thereby using 
Broyden's algorithm\cite{Broyden} for better convergence.
\begin{figure}
\includegraphics[width=0.9\columnwidth]{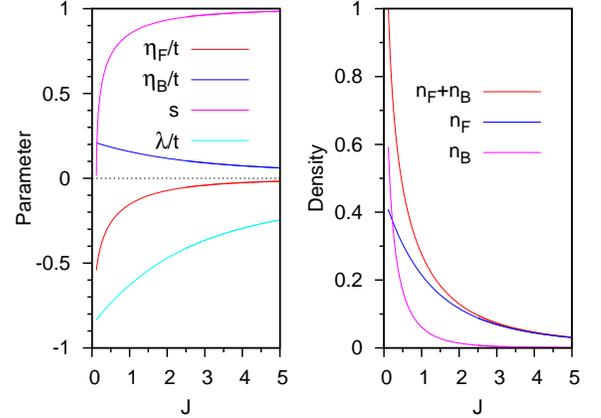}
\caption{\label{fig1} 
Self-consistent parameters of the mean-field solution versus $J/t$ (Left).
Densities of the fermions and bosons versus $J/t$ (Right).}
\end{figure}
For the Kondo-insulator with nearest-neighbor hopping particle hole-symmetry
results in $\zeta_{\bm k}=\tilde{\zeta}_{\bm k}=0$. Figure \ref{fig1} shows the
remaining parameters, $\eta$, $s$ and $\lambda$ as functions
of $J/t$. The parameter $s$ reaches zero for $J_{min}/t\approx 0.1173$ 
and there is no solution for smaller values of $J/t$.
The reason is that even for $s=0$ the parameters $\Delta_{\bm k}$ and
$\tilde{\eta}$ are finite and in fact
increase for small $J/t$ so that the resulting density of particles,
$n_B+n_F$, exceeds $1$ at $J_{min}$ and (\ref{constraint_1})
can no longer be fulfilled. This may be a consequence of the
fact that the bond particle formulation of the Kondo lattice ultimately
is a strong-coupling theory which is justified best for 
$J/t\rightarrow \infty$. It should also be noted that once 
the particle density $n_B+n_F$ approaches unity, the bond particle theory is 
highly unreliable anyway. \\
The mean-field expectation values $\eta_{\bm k}$ and 
$\tilde{\eta}_{\bm k}$ are small for $J/t>1$,
the parameter $\lambda$ is relatively large and negative. 
\begin{figure}
\includegraphics[width=0.9\columnwidth]{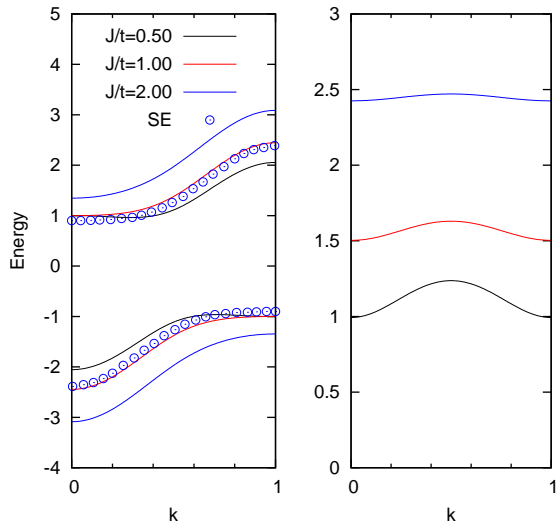}
\caption{\label{fig2} 
Quasiparticle bands (Left) and triplet frequency (Right) for the 1D Kondo
insulator from mean field theory, wave vectors in unit of $\pi$.
Data points
were obtained by SE for $J/t=2$\cite{seriesexp}.}
\end{figure}
Figure \ref{fig2} shows the bands $E_{\nu,{\bm k}}$
for the fermions and the dispersion $\omega_{{\bm k}}$ of the bosons.
The smallness of $\tilde{\eta}$ results in a small bandwidth for the bosons,
whereas the relatively large and negative $\lambda$ results in a large
bandgap for the fermions which stays approximately constant
for $J/t<1$, as well as a considerable upward shift of the triplet dispersion. 
The band structure is consisten with the hybridization picture
with extended `heavy' band portions and is roughly consistent with numerical 
results results for the 1D PAM\cite{Tsutsui,Carsten} and KLM\cite{seriesexp} 
although the size of the gap comes out too large.
QMC has also shown a well-defined weakly dispersive and gapped mode in the 
dynamical spin correlation function of the PAM\cite{Carsten}, roughly 
consistent with the mean-field boson dispersion. 
The boson dispersion is symmetric with 
respect to $k=\frac{\pi}{2}$ whereas DMRG calculations find the maximum of the 
dispersion of the lowest triplet state at $k=0$, the minimum 
at $k=\pi$\cite{yuwhite}. We define the quasiparticle gap, 
$\Delta_{QP}=E_0^{(N+1)}+E_0^{(N-1)} - 2E_0^{(N)}$ which we approximate by the band 
gap i.e. in 1D $\Delta_{QP}=E_\beta(k=0)-E_\alpha(k=\pi)$.\\
So far we have ignored the terms $H_3$ and $H_4$ because a mean-field treatment
of these terms would result in some type of magnetic order.
To study the contribution of $H_3$ and $H_4$
as well as the unfactorized remainder of $H_2$ to the ground state energy 
at least approximately, we treat these terms in 2$^{nd}$ order perturbation 
theory in analogy to M{\o}ller-Plesset perturbation theory\cite{MP}.
Since the Kondo insulator has a finite gap in its excitation spectrum
this is probably a good approximation.
More precisely, we take the mean-field Hamiltonian
$H_{MF}=H_F+H_B+Nc$ as the unperturbed Hamiltonian -
its ground state is the product
of the ground states of $H_F$ and $H_B$. The perturbation is
\begin{eqnarray*}
\tilde{H}_1 &=&  H_2 + H_3 + H_4 - (H_{2,MF} - N\Theta)
\end{eqnarray*}
where $H_{2,MF}$ is the mean-field factorized
form of $H_2$ i.e. the terms in $H_F+H_B$ which are
$\propto \eta, \zeta, \tilde{\eta}, \tilde{\zeta}$.
$\tilde{H}_1$ thus is a sum of terms of the form
\begin{eqnarray*}
\sum_{m,n}\frac{t_{m,n}}{2}&&\left(\;O_{m,n}^{(F)}\;O_{m,n}^{(B)} 
- O_{m,n}^{(F)}\; \langle O_{m,n}^{(B)} \rangle \right. \nonumber \\
&& \left. \;\;\;\;\;\;\;\;- \langle O_{m,n}^{(F)} \rangle\;O_{m,n}^{(B)}
+\langle O_{m,n}^{(F)} \rangle\; \langle O_{m,n}^{(B)} \rangle \;\right)
\end{eqnarray*}
where $O_{m,n}^{(F)}$ ($O_{m,n}^{(B)}$)
contain only fermion (boson) operators and $\langle ..\rangle$
denotes expectation values in the mean-field ground state
(which are zero for terms arising from $H_3$ and $H_4$). Then 
$H_{MF}+\tilde{H}_1$ is the complete Hamiltonian (with added constraints)
and the first order correction $\langle\tilde{H}_1\rangle=0$.  
All matrix elements of $\tilde{H}_1$ between the mean-field ground state
$|GS\rangle$ and states which contain either only a fermionic excitation - 
such as $\beta_{{\bm k},\sigma}^\dagger \alpha_{{\bm k},\sigma'}^{}|GS\rangle$ - 
or only a bosonic excitation - such as
$\tau_{{\bm q},x}^\dagger \tau_{-{\bm q},y}^\dagger|GS\rangle$ - are zero.
It follows that in 2$^{nd}$ order perturbation theory we may as well
take the perturbation to be $\tilde{H}_1=H_2+H_3+H_4$
but consider only intermediate states which contain both, 
a fermionic {\em and} a bosonic excitation.\\
We begin with $H_3$, which can be rewritten as
\begin{eqnarray*}
H_3 &=& -\sum_n \;(\frac{t_{m,n}}{2}\;{\bm t}^\dagger_{n}\cdot {\bm A}_n  + H.c.)\\
{\bm A}_n &=&\sum_m\;\left[\;({\bm b}_{m,n}-{\bm a}_{m,n})\;s_m +
s_m^\dagger\;({\bm \pi}_{m,n} - {\bm \pi}_{n,m})\;\right].
\end{eqnarray*}
A considerable simplification comes about by noting that
since the mean-field expectation value $\tilde{\eta}$ is small,
resulting in an almost flat triplet dispersion,
$\omega_{\bf k}\approx \tilde{J}$, 
we may neglect all terms in the triplet Hamiltonian 
other than the energy term 
$\tilde{J}\sum_n\;{\bf t}_n^\dagger \cdot {\bf t}_n^{}$.
The ground state then is the vacuum for triplets and only
the terms $\propto {\bm t}^\dagger_n$ contribute to the energy correction.
In the Kondo insulator the operators ${\bm A}_n$, being quadratic in the 
fermions, can only excite a quasiparticle from the lower to the upper 
quasiparticle band, say from momentum ${\bm p}$ to momentum ${\bm q}$.
A typical state which couples to the ground state in this way
would be form $\beta_{{\bm q},\sigma}^\dagger \alpha_{{\bm p},\sigma}^{} 
{\bm t}_{n,z}|GS\rangle$.
\begin{figure}
\includegraphics[width=0.5\columnwidth]{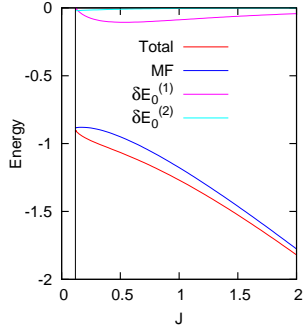}
\caption{\label{fig3} 
Contributions to the ground state energy per site: mean-field energy and
M{\o}ller-Plesset correction. The vertical line denotes $J_{min}/t$ where
$s\rightarrow 0$.}
\end{figure}
The unperturbed energy of this state is 
$E_{\beta,{\bm q}} + \tilde{J} - E_{\alpha,{\bm p}} + E_0$,
because the triplet which is created along with the particle-hole
pair contributes the energy $\tilde{J}$ if we neglect the dispersion of the 
triplets. The matrix element for the transition can be evaluated
by using (\ref{unitary}) and is 
\begin{eqnarray*}
\frac{s}{2N}\;e^{i({\bm p}-{\bm q})\cdot {\bm R}_n}\;m_{{\bm p},{\bm q}}
\end{eqnarray*}
with
\begin{eqnarray*}
m_{{\bm p},{\bm q}}&=&
\epsilon_{\bm q} u_{\bm p} (u_{\bm q}+v_{\bm q}) -
\epsilon_{\bm p} u_{\bm q} (u_{\bm p}-v_{\bm p}).
\end{eqnarray*}
The correction to the energy/site due to creation of a single
triplet then is
\begin{eqnarray}
\delta E_0^{(1)} &=& -\frac{3s^2}{2N^2}\;\sum_{{\bm p},{\bm q}}\;
\frac{m_{{\bm p},{\bm q}}^2}{E_{\beta,{\bm q}} + \tilde{J} - E_{\alpha,{\bm p}}}.
\label{de2}
\end{eqnarray}
We proceed to the correction due to $H_2$ and $H_4$. The parts
which give a nonvanishing result when acting onto the vacuum for triplets are
\begin{eqnarray*}
H_2' &=&\;\sum_{m,n}\;\frac{t_{m,n}}{2}\left(\;a_{n,\downarrow}b_{m,\uparrow} -
a_{n,\uparrow}b_{m,\downarrow}\;\right)\;{\bm t}^\dagger_{m} \cdot {\bm t}^\dagger_{n}
\nonumber \\
H_4' &=& -i\;\sum_{m<n} \;\frac{t_{m,n}}{2}
\left(\;{\bm \pi}_{m,n} - {\bm \pi}_{n,m}\;\right)
\cdot ({\bm t}^\dagger_{m} \times {\bm t}^\dagger_{n} )\nonumber \\
\end{eqnarray*}
The states which can be reached have the form
$\beta_{{\bm q},\sigma}^\dagger \alpha_{{\bm p},\sigma'}^{} 
{\bm t}_{n,x}^\dagger {\bm t}_{m,x'}^\dagger|GS\rangle$.
Proceeding as above and
specializing to a 1D chain with only nearest neighbor hopping $-t$
we find for the energy shift due to the creation of two
triplets
\begin{eqnarray}
\delta E_0^{(2)} &=&-\frac{9t^2}{N^2}\;\sum_{{\bm p},{\bm q}}\;
\frac{(1 + \cos({\bm p}+{\bm q}))\;u_{\bm p}^2 u_{\bm q}^2}
{E_{\beta,{\bm q}} + 2\tilde{J} - E_{\alpha,{\bm p}}}.
\label{de3}
\end{eqnarray}
The different contributions to the ground state energy are shown in
Figure \ref{fig3}.
The perturbation correction is quite small which indicates that the use of
perturbation theory is adequate. 
The energy shift due to creation of a single triplet, 
$\delta E_0^{(2)}$ is small but still of order $0.1t$, 
whereas $\delta E_0^{(3)}$ is negligible, of order $10^{-2}t$.\\
\begin{figure}
\includegraphics[width=0.8\columnwidth]{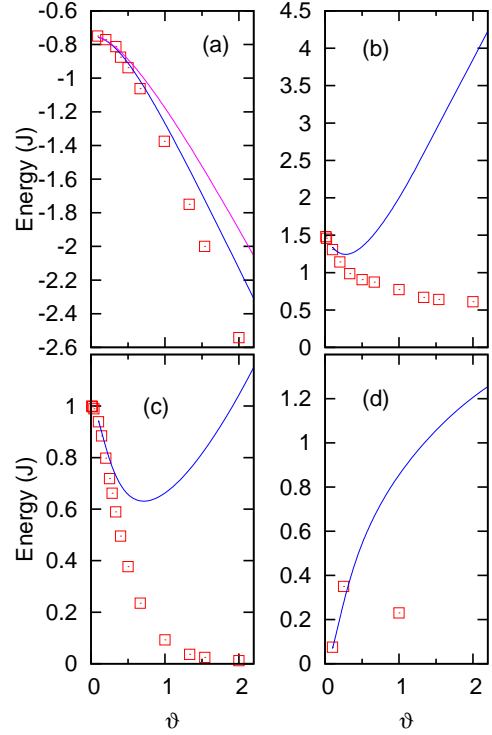}
\caption{\label{fig4} 
Characteristic energies for the 1D Kondo-insulator
by bond-particle mean-field theory (blue lines) compared to 
DMRG\cite{yuwhite} (red squares): ground state energy/site (a),
quasiparticle gap $\Delta_{QP}$ (b), spin gap $\Delta_{S}$ (c)
and spin excitation bandwidth $W_s$ (d) versus $J/t$. The magenta 
line in (a) is the mean-field energy without M{\o}ller-Plesset correction,
$\vartheta=t/J$}
\end{figure}
Lastly, we discuss an improved calculation of the triplet dispersion
$\omega_{\bf k}$. While the mean-field calculation predicts these to
be almost dispersionless the term $H_3$ gives a more substantial dispersion.
We make the variational ansatz for a $z$-triplet-like excitation
with momentum ${\bf q}$
\begin{eqnarray}
|\Psi_{\bf q}\rangle &=&
\left[\;a_{\bf q} {\bf t}_{{\bf q}}^\dagger + \frac{1}{\sqrt{2N}}
\sum_{{\bm k},\sigma,\sigma'}
b_{{\bf q},{\bf k}}\;\beta_{{\bf k}+{\bf q},\sigma}^\dagger 
{\bm \tau}_{\sigma,\sigma'}
\alpha_{{\bf k},\sigma'}^{}\;\right]|0\rangle,\nonumber \\
\label{svar}
\end{eqnarray}
with variational parameters $a_{\bf q}$ and $b_{{\bf q},{\bf k}}$.
We obtain the triplet frequency
\begin{eqnarray}
\tilde{\omega}_{\bm q} &=&\tilde{J} +\frac{s^2}{2N}\;\sum_{\bm k}\;
\frac{V^2({\bm k},\bm{q})}{\tilde{\omega}_{\bm q}-(E_{\beta,{\bm k}+\bm{q}} -
E_{\alpha,{\bm k}})}\nonumber \\
V({\bm k},\bm{q})&=&
(\epsilon_{{\bm k}+\bm{q}}-\epsilon_{{\bm k}})\;v_{{\bm k}+\bm{q}} v_{{\bm k}}\nonumber \\
&&\;\;\;\;\;\;\;\;+ \epsilon_{{\bm k}+\bm{q}}\;u_{{\bm k}+\bm{q}} v_{{\bm k}}
+ \epsilon_{{\bm k}}\;u_{{\bm k}}v_{{\bm k}+\bm{q}},
\label{etrip}
\end{eqnarray}
where $\epsilon_{{\bm k}}$ is the noninteracting dispersion of the
conduction electrons.
Numerical evaluation shows that this always takes its minimum
at $k=\pi$ and the maximum at $k=0$ - consistent with
DMRG\cite{yuwhite}. The energy $\tilde{\omega}_{q=\pi}$ thus
is the energy of the lowest $S=1$ excitation, called the spin gap, 
$\Delta_{s}$. The bandwidth of the spin excitations is 
$W_s=\tilde{\omega}_{q=0}-\tilde{\omega}_{q=\pi}$.
Figure \ref{fig4} compares the dependence of the ground state energy per site,
the quasiparticle and spin gap and the bandwidth of the spin excitations
on $t/J$ for the 1D Kondo insulator to results obtained by DMRG\cite{yuwhite}. 
As expected, the perturbation correction improves the 
ground state energy/site, which is reasonably close to the numerical values
for $t/J<1$. For all other quantities the calculated energies
deviate from the numerical results already for relatively large $J/t$. 
One might 
wonder if this is the consequence of the additional approximation
to neglect $\tilde{\eta}$, but even for $J/t=1$ where the boson bandwidth
is quite small (see Figure \ref{fig2}) the deviation for the
spin gap is already substantial.
To conclude this section we discuss the differences to the previous mean-field treatment 
in Ref.\cite{JureckaBrenig}. In this work, both the singlet and
the triplet operators in (\ref{hamtot}) were replaced by $c$-numbers,
$s_n^\dagger,s_n^{} \rightarrow s$ and $\bm{t}_n^\dagger,\bm{t}_n^{} \rightarrow \bm{t}\;e^{i\bm{Q}\cdot\bm{R}_n}$, thus reducing (\ref{hamtot}) to a quadratic form
from the outset,
and then minimizing $F$ with respect to $s$ and $\bm{t}$.
The dynamics of the spin excitations thereby was not studied.
\section{Renormalized Energy of Formation}
We consider a different approximation scheme to account for
the constraint whereby we consider sites occupied by
a singlet as `empty'. This is equivalent to working in a fictitious
Hilbert space for the fermionic particles $a^\dagger_{n,\sigma}$ and
$b^\dagger_{n,\sigma}$ as well as the triplets ${\bf t}^\dagger_{n,\alpha}$,
whereby the states in this fictitious Hilbert space correspond to those of 
the physical Kondo lattice according to the rule
\begin{eqnarray}
\prod_{i\in S_a}a_{i,\sigma_i}^\dagger\;\prod_{j\in S_b}b_{j,\sigma_j}^\dagger\;
\prod_{l\in S_t}t_{l,x_l}^\dagger|0\rangle \rightarrow 
\;\;\;\;\;\;\;\;\;\;\;\;\;\;\;\;\;\;\;\;\;\;\;\;\;\;\;
\nonumber \\
\bigotimes_{i\in S_a}|a,i,\sigma_i\rangle \;\bigotimes_{j\in S_b}|b,j,\sigma_j\rangle
\;\bigotimes_{l\in S_t}|t_{x_l,l}\rangle\;
\bigotimes_{n\in S_s}\;|s_n\rangle.
\label{trans}
\end{eqnarray}
$S_a$, $S_b$ and $S_t$ denote the set of sites occupied by a hole-like fermion,
an electron-like fermion or a triplet, respectively, and
$S_s=(S_a\cup S_b\cup S_t)^C$ the set of remaining sites.
In other words, all sites not occupied by a fermion or triplet
are filled up with `inert' singlets. The Hamiltonian - and all other operators 
in the bond particle representation - then can be obtained from
(\ref{hamtot}) by replacing all singlet operators by unity. 
Only the form (\ref{HJ}) of the exchange term can be used.
For this representation to
make sense we again have to impose the constraint that no two particles
of any type occupy the same site, because the resulting state could not
be translated meaningfully to a state of the physical Kondo lattice
via (\ref{trans}). Assuming that the density of bond particles is small,
however, we relax again this constraint. \\
On the other hand, if the constraint were rigorously enforced, presence of 
any one particle -  be it $a^\dagger_{n,\sigma}$, $b^\dagger_{n,\sigma}$ or 
${\bf t}^\dagger_{n,\alpha}$ - at a given site $n$ would prevent all remaining 
terms in the Hamiltonian which involve creation or annihilation of any other 
particle at site $n$ from acting, resulting in a loss of kinetic
energy. The constraint 
thus increases the cost in energy for adding a fermion or boson.
Accordingly, in (\ref{HJ}) the energy for adding a fermion therefore should be
$e_0=3J/4+\kappa$ rather than $3J/4$ and the energy for adding
a boson $\tilde{J}=J+\kappa$ rather than $J$, where $\kappa$ is some as 
yet unspecified loss of kinetic energy.
Actually $\kappa$ may be expected to be different for fermions and bosons.
We will
discuss possible estimates for $\kappa$ later on. It should
also be noted that such an increase of the 
energies of formation of the particles would reduce their densities and thus 
make relaxing the constraint of no double occupancy an even better 
approximation.
The mean-field theory outlined in the previous section and the
approximation scheme discussed in the present section mimick the constraint
in different ways: mean-field theory amounts to a 
Gutzwiller-like downward renormalization
of the hopping integrals whereas the present scheme amounts to a
higher energy of formation of the particles.
Collecting all terms which become quadratic when we drop the singlets
we obtain the noninteracting Hamiltonian
\begin{widetext}
\begin{eqnarray*}
H_0&=&\sum_n\;\left(\;e_0\;\sum_\sigma \;(b_{n,\sigma}^\dagger b_{n,\sigma}^{} 
+ a_{n,\sigma}^\dagger a_{n,\sigma}^{}) + \tilde{J}\;\bm{t}_n^\dagger \cdot \bm{t}_n^{}\;\right) +\sum_{m,n}\;\frac{t_{m,n}}{2}\;\sum_\sigma \;
\left(\;b_{m,\sigma}^\dagger b_{n,\sigma}^{} -
a_{m,\sigma}^\dagger a_{n,\sigma}^{}\;\right)\nonumber \\
&&\;\;\;\;
-\sum_{m,n}\;\frac{t_{m,n}}{2}\; 
\left[\;\left(\;
b_{m,\uparrow}^\dagger a_{n,\downarrow}^\dagger
- b_{m,\downarrow}^\dagger a_{n,\uparrow}^\dagger
\;\right) + H.c.\;\right] - \frac{3NJ}{4}.
\end{eqnarray*}
The interaction part of the Hamiltonian is the sum of
\begin{eqnarray}
H_2&=&\sum_{m,n}\;\frac{t_{m,n}}{2}\;\sum_\sigma \;\left(\;b_{m,\sigma}^\dagger b_{n,\sigma}^{} -
a_{m,\sigma}^\dagger a_{n,\sigma}^{}\;\right)\;\bm{t}_n^\dagger\cdot \bm{t}_m^{}\;
+\sum_{m,n}\;\frac{t_{m,n}}{2}\;
\left[\;\left(\;
b_{m,\uparrow}^\dagger a_{n,\downarrow}^\dagger
- b_{m,\downarrow}^\dagger a_{n,\uparrow}^\dagger
\;\right) \bm{t}_m^{}\cdot \bm{t}_n^{}\;+ H.c\;\right]\nonumber\\
H_3&=& -\;\sum_{m,n}\;\frac{t_{m,n}}{2}\;\left[\;\left(\;{\bm \pi}^\dagger_{m,n}\cdot
\left(\;\bm{t}_n^{} - \bm{t}_m^{}\right) + H.c.\;\right)
+\left(\;\bm{b}_{m,n} - \bm{a}_{m,n}\;\right)\cdot
(\bm{t}_n^\dagger + \bm{t}_m^{}\;)\;\right],\nonumber\\
H_4&=&
\sum_{m,n}\;\frac{t_{m,n}}{2}\;\left[\;\left[\;i{\bm \pi}^\dagger_{m,n}\cdot
(\;\bm{t}_m^{}\times\bm{t}_n^{}\;)
+ H. c. \;\right]
- i\;(\;\bm{b}_{m,n} - \bm{a}_{m,n}\;)\;
(\;\bm{t}_n^\dagger\times \bm{t}_m^{}\;)\;\right].
\label{hamtot2}
\end{eqnarray}
\end{widetext}
The part $H_0$ was used in Refs.\cite{Oana,afbf}.
Due to particle-hole symmetry the extra Lagrange multiplier introduced 
in these Refs. to enforce consistency of
the $c$-like spectral weight with $N_e$ is not necessary here.
We now proceed as in the case of mean-field theory, that means
first diagonalize $H_0$ to obtain the band structure and treat the interaction 
terms in the same approximation as there, i.e. in perturbation theory for the 
ground state energy and using the variational ansatz (\ref{svar}) for the spin 
excitations.
The fermionic part $H_F$ again can be diagonalized by the unitary 
transformation (\ref{unitary}) with the result
\begin{eqnarray*}
H_F&=&\sum_{\bm{k},\sigma}\;(\;E_{\alpha,\bm{k}}\; \alpha_{\bm{k},\sigma}^\dagger \alpha_{\bm{k},\sigma}^{}
+ E_{\beta,\bm{k}}\; \beta_{\bm{k},\sigma}^\dagger \beta_{\bm{k},\sigma}^{}\;)\\
&& \;\;\;\;\;\;\;\;\;\;\;\;\;\;\;\;\;\;\;\;\;\;\;\;\;\;\;\;\;\;\;\;
-\;2\sum_{\bm{k}} \;(\frac{\epsilon_{\bm{k}}}{2} - e_0), \\
E_{\nu,\bm{k}}&=& \frac{\epsilon_{\bm{k}}}{2} \pm W_{\bm{k}} - \mu,\\
W_{\bm{k}}&=&\sqrt{ e_0^2 + \left(\frac{\epsilon_{\bm{k}}}{2}\right)^2},\\
u_{\bm{k}}&=&-\sqrt{\frac{W_{\bm{k}}-e_0}{2W_{\bm{k}}}},\\
v_{\bm{k}}&=&\frac{-\frac{\epsilon_{\bm{k}}}{2}}{\sqrt{2W_{\bm{k}}(W_{\bm{k}}-e_0)}}.
\end{eqnarray*}
The noninteracting ground state for the bosons is the bosonic
vacuum i.e. the bosons do not contribute to the
ground state energy $E_0$ in this approximation.
The Helmholtz Free energy is
\begin{eqnarray*}
F &=& -\frac{2}{\beta}\;\sum_{\bm{k}}\;\sum_{\nu\in\{\alpha,\beta\}}\;
\log\left(1+e^{-\beta E_{\nu,\bm{k}}}\right) \nonumber \\
&& \;\;\;\;\;\;\;\;+ N(2e_0 -\frac{3J}{4}) + \mu N.
\end{eqnarray*}
We can obtain the expectation value of the kinetic energy
by multiplying all hopping integrals by a parameter $\chi$: 
$t_{m,n}\rightarrow \chi t_{m,n}$ and forming 
$\frac{\partial F}{\partial \chi}|_{\chi=1}$, with the result:
\begin{eqnarray}
k&=&\frac{1}{N}\;\sum_{{\bm k}}\;\left[\;\epsilon_{{\bm k}}
\left(\;1-\frac{\epsilon_{{\bm k}}}{2W_{{\bm k}}}\;\right) f(E_{\alpha,{\bm k}})
\right.
\nonumber \\
&&\left. \;\;\;\;\;\;\;\;\;\;\;\;\;\;\;\;+ \epsilon_{{\bm k}}
\left(\;1+\frac{\epsilon_{{\bm k}}}{2W_{{\bm k}}}\;\right) f(E_{\beta,{\bm k}})\right].
\label{kin}
\end{eqnarray}
Treating the interaction part $H_2+H_3+H_4$ in 2$^{nd}$ order perturbation theory
a slight modification occurs. Since the ground state is the vacuum for
bosons and the vector operators ${\bf a}$, ${\bf b}$ and ${\bf \pi}$
have zero expectation value in the fermionic ground state (which is spin 
singlet) only intermediate states which contain both, a bosonic and a 
fermionic excitation, can be reached from the ground state by acting with
$H_3$ or $H_4$. The energy shifts due to such doubly excited intermediate
states take the form (\ref{de2}), (\ref{de3}), but with $s\rightarrow 1$ 
in (\ref{de2}). In addition, however, the term $H_2$ also has a
nonvanishing matrix element with states of the form
${\bf t}_{m}^\dagger \cdot {\bf t}_{n}^\dagger |GS\rangle$, because the
fermionic factor of the corresponding term is a singlet, which does have
a nonvanishing ground state expectation value.
For the 1D chain with nearest neighbor hopping
this gives an additional contribution to
$\delta E_0^{(2)}$ of $-3\;\frac{I^2}{2\tilde{J}}$, whereby
\begin{eqnarray}
I &=&-\frac{t^2}{N}\;\sum_q\;\frac{\cos^2(q)}{W_q}.
\label{bigshift}
\end{eqnarray}
Numerical evaluation shows that this contribution is quite substantial
and obviously this replaces the energy gain due to
the mean-field factorized terms in the previous section.
Finally, the equation for the dispersion of the spin excitation takes the
form (\ref{etrip}), again with $s\rightarrow 1$.\\
\begin{figure}
\includegraphics[width=0.6\columnwidth]{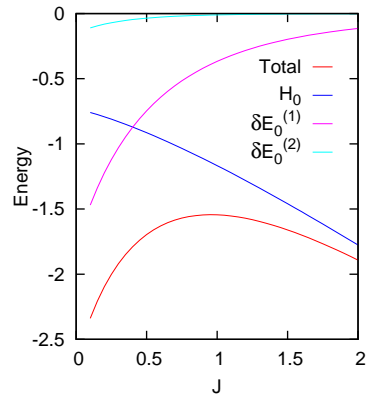}
\caption{\label{fig5} 
Contributions to the ground state energy per site: Contribution from $H_0$
and corrections from perturbation theory.}
\end{figure}
Lastly, we consider the value of $\kappa$,
the correction to the energies of formation of the fermions and bosons.
\begin{figure}
\includegraphics[width=0.8\columnwidth]{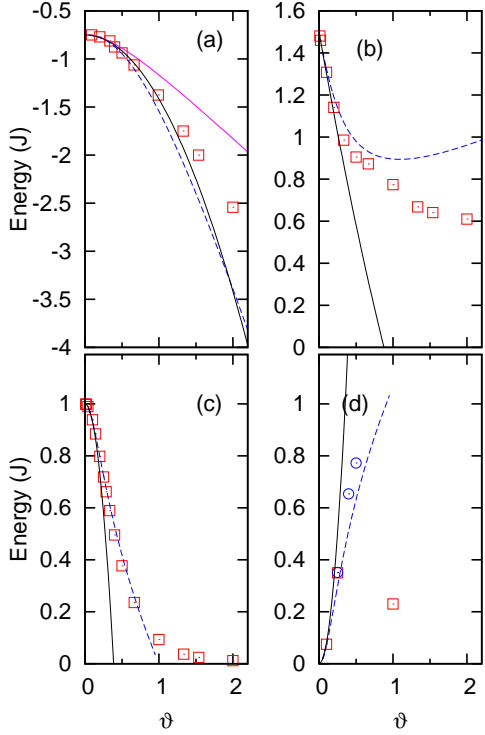}
\caption{\label{fig6} 
Characteristic energies in  the 1D Kondo-insulator from the renormalized 
energy of formation scheme with $x=0.4$ (blue lines) compared to 
DMRG\cite{yuwhite} (red squares) and SE\cite{seriesexp} (blue circles):
ground state energy/site (a), quasiparticle gap $\Delta_{QP}$ (b), 
spin gap $\Delta_{S}$ (c) and spin excitation bandwidth $W_s$ (d) versus $J/t$. 
The magenta line in (a) is the energy from $H_0$ without perturbation theory 
correction. The black lines are obtained by perturbation expansion
in $t/J$\cite{uedareview}, $\vartheta=t/J$.}
\end{figure}
\begin{figure}
\includegraphics[width=0.8\columnwidth]{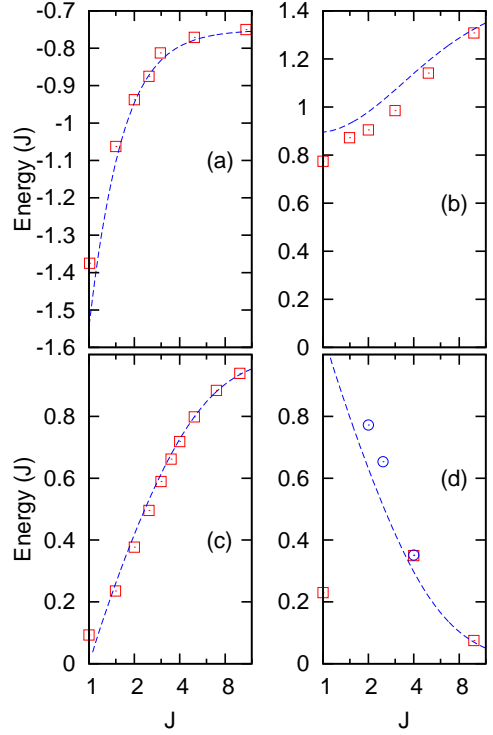}
\caption{\label{fig7} 
Same as Figure \ref{fig6} but all energies plotted versus $J/t$ in the range
$J/t\ge 1$.}
\end{figure}
We switch to a phenomenological approach and
approximate $\kappa \approx x k$, i.e. a dimensionless
parameter $x$ times the kinetic energies of the Fermions/site, given
in (\ref{kin}). With $x$ fixed, 
$\kappa$ has to be determined self-consistently for each $J/t$. 
We neglect the loss of kinetic energy of bosons which is 
reasonable for $J/t>1$ where the boson density is low. In fact, as will be
shown in a moment, we can obtain good agreement with numerics over the whole 
range $J/t>1$ by choosing $x\approx 0.4$ independent of $JJ/t$.
Varying $0.2<x<0.6$ thereby does not deteriorate the 
agreement significantly, so that also an $x$ which varies with $J/t$ - which
is actually what one might expect - would give similar results.\\
To begin with, Figure \ref{fig5} shows the various contributions to the ground 
state  energy/site obtained with this choice of $x$ and demonstrates that at 
least for $J/t>1$ the perturbation correction is small as it should be. 
Figure \ref{fig6} compares characteristic energies of the system as functions 
of $t/J$ to numerical results. For larger $t/J>1$ the agreement is
poor, in particular for $t/J \approx 1$ the spin excitation energy 
$\tilde{\omega}_q$ from the variational ansatz becomes negative around 
$q=\pi$ indicating the failure of the calculation. Accordingly, results for 
the spin gap $\Delta_s$ and spin excitation bandwidth $W_s$ are shown only up 
to this value of $t/J$.
$W_s$ comes out quite good for $t/J< 1$ - the Figure also shows the
very good agreement between DMRG and  SE for $W_s$ at $t/J=0.25$ in (d).
DMRG and SE also agree very well for $\Delta_s$ so that we do not show 
the SE results in (c). Figure \ref{fig6} also shows results obtained by
perturbation expansion in $t/J$\cite{uedareview}. 
As one might have expected the DMRG results and bond particle theory approach 
these for $t/J\rightarrow 0$. The ground state energy is reproduced
remarkably well be the perturbation expansion, but all other
characteristic energies deviate substantially from the
perturbation expansion for $t/J\rightarrow 1$. This shows that despite
being a strong coupling theory by nature, bond
particle theory does go beyond simple perturbation theory.
Figure \ref{fig7} shows the same characteristic
energies but now plotted versus $J/t$ in the range $J/t>1$. It is obvious
that in this range the agreement between bond particle theory
and numerics is quite good.
Figure \ref{fig8} shows the densities of fermions and bosons,
$n_F$ and $n_B$. Thereby $n_B$ is
obtained from $n_B=\frac{\partial E_0}{\partial \tilde{J}}$ where
$E_0$ is the ground state energy per site
including the second order perturbation 
correction. The main contribution thereby comes from 
(\ref{bigshift}). The data points for $t/J=1$ are DMRG 
results\cite{yuwhite}. The densities are small for $t/J<1$ and bond 
particle theory somewhat underestimates the densities of the particles
at $t/J=1$. For larger $t/J$ the densities increase rapidly and
the sum $n_B+n_F$ exceeds $1$ at $t/J\approx 2$, indicating the
breakdown of bond particle theory.
Figure \ref{fig9} shows the band structure of the fermions and the
dispersion of the bosons obtained from the variational ansatz.
For $J/t=2$ the dispersions can be compared to results obtained
by SE taken from Ref. \cite{seriesexp}.
While the quasiparticle gap is approximately correct,
the `heavy' part of the band structure has too much dispersion
as compared to SE, and the bandwidth is slightly overestimated.
The dispersion of the spin excitations is reasonably correct
for $J/t=2$, both the spingap, the overall form of the dispersion
and the bandwidth compare quite well with the SE result.
\begin{figure}
\includegraphics[width=0.6\columnwidth]{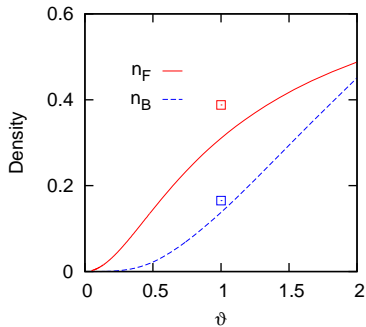}
\caption{\label{fig8} 
Densities of fermions and bosons for the 1D Kondo insulator versus $\theta=t/J$,
data points from DMRG\cite{yuwhite}.}
\end{figure}
\begin{figure}
\includegraphics[width=0.8\columnwidth]{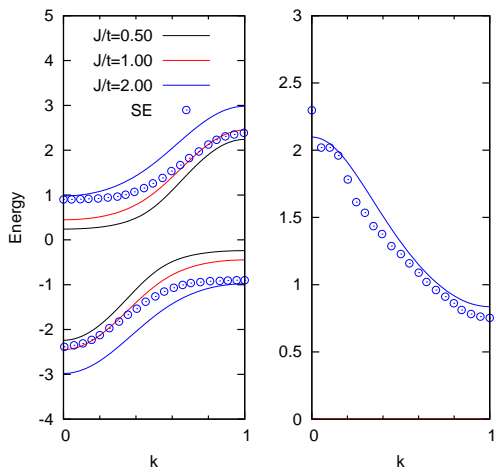}
\caption{\label{fig9} 
Band structure of the fermions (left) and boson dispersion from the 
variational ansatz (\ref{etrip}) (right) for different $J/t$,
wave vectors in units of $\pi$. Data points
obtained by SE for $J/t=2$\cite{seriesexp}. The boson
dispersion for $J/t=1$ and $J/t=0.5$ is not given because there the variational
calculation gives negative energy over some range of $k$.}
\end{figure}
The combined DMRG and SE data in Figure \ref{fig6} suggest
that there is a crossover between two regimes at around $t/J\approx 1$: 
the $\Delta_s/J$ vs. $J/t$ curve drops rapidly for $J/t>1$ but then bends 
sharply at $J/t\approx 1$ and $\Delta_s/J$ is small but finite for $J/t<1$. 
Similarly, the band width of the spin excitations, $W_s/J$ increases
with decreasing $J/t$ in the range $J/t>1$ but then must drop sharply at 
$J/t\approx 1$. This may indicate a crossover from a strong coupling
regime for $J/t>1$ where the system apparently can be described well
by the bond particle theory, to a weak coupling regime for $J/t<1$
where maybe mean-field theories work better.
It should also be noted that the present theory {\em must} fail in the limit 
$J/t\rightarrow 0$ not only because the bond particle density increases
sharply but also because for because $J/t\rightarrow 0$ the quasiparticle gap 
vanishes, whereas the energy $e_0$ - which determines the 
magnitude of the gap - cannot approach zero for any $x>0$.
All in all Figure \ref{fig6} indicates that for $t/J<1$ the bond-particle 
Hamiltonian (\ref{hamtot2}) with suitably renormalized
energies of formation gives a reasonably correct description of the
low-energy elementary excitations of the 1D Kondo insulator.
\section{Summary and Discussion}
In summary we have derived an exact representation of the Kondo lattice
model in terms of bond particles: fermions corresponding to
unit cells with an odd number of electrons and bosons corresponding to
cells with two electrons coupled to a singlet or triplet.
Thereby the constraint to have
precisely one $f$ electron/cell, which considerably complicates the solution of
the KLM, is fulfilled automatically and replaced by
the constraint to have precisely one bond particle per site.
If the singlet bosons are considered as condensed or the singlet is
defined as the vaccum state of a cell, this constraint becomes an infinitely 
strong Hubbard-like repulsion, but for a relatively dilute system of particles, 
so that it may be justified to relax it
(for a system of low density even an infinitely strong repulsion can be treated 
diagrammatically\cite{FW,Kotov,Shevchenko}).
The requirement of low particle density is indeed fulfilled for $J/t>1$.\\
We have discussed two schemes to approximately
incorporate effects of the remaining Hubbard repulsion between bond particles 
into their Hamiltonian. First, mean-field theory, where the singlets are taken 
as condensed, amounts to a Gutzwiller-like downward renormalization of 
all hopping integrals. Second, a scheme where the singlet is considered as
the vacuum state of a site and the constraint is mimicked by adding the 
loss of kinetic energy, which incurs due to the 
blocking of a site by a bond particle, to the energy ascribed
to the respective particle. Approximating this loss of kinetic energy as 
the kinetic energy of fermions per site times a phenomenological constant
of order unity allowed to reproduce numerical results ontained by 
density matrix renormalization group
and series expansion calculations for the 1D Kondo insulator in the range 
$J/t>1$ with good accuray.
Thereby relatively simple techniques were used
- 2$^{nd}$ order perturbation theory for the ground state energy and
the simplest possible variational wave function for the triplet dispersion -
to produce these results. 
The good agrement with numerics in the range $J/t>1$ for a variety of
quantitiesis then is a strong indication that in this parameter range 
the triplets and fermions indeed correspond to the approximate elementary 
excitations and this is the main result of the present paper.
Despite being a somewhat lengthy
expression the bond particle Hamiltonian with renormalized particle
energies appears to be useful for quantitative calculations.
Of course the phenomenological approach used here is somewhat unsatisfactory
and a more rigorous calculation following Refs. \cite{Kotov,Shevchenko}
would be desirable.\\
The question then arises, whether $J/t>1$ is a sufficient range of validity
to discuss magnetic ordering and quantum critical points in the KLM.
For the 2D square lattice with nearest neighbor hopping it is
known that at $n_e=2$ and $T=0$ antiferromagnetic ordering occurs for
$J/t \le J_c/t=1.45$\cite{Assaad}, i.e. for relatively large $J/t$
(although with the information at hand we cannot say much about the range 
of vailidity of the bond particle description in higher dimensions).
As pointed out by Sachdev and Bhatt\cite{SachdevBhatt} bond particle theory 
gives a rather natural description of magnetic ordering, namely
the condensation of triplets into a momentum corresponding
to the magnetic ordering wave vector. Applying this description
of antiferromagnetic ordering in the KLM has already produced encouraging
results: using the mean-field version of bond particle theory
with condensed triplets, Jurecka and Brenig found $J_c/t=1.5$, quite
close to the exact value. This is even more remarkable in that even 
numerical methods appear to have difficulties to accurately reproduce
$J_c/t$: VMC gives $J_{c,1}/t=1.7$\cite{WatanabeOgata}, DCA gives 
$J_{c,1}/t=2.1$\cite{MartinBerxAssaad}, and DMFT gives 
$J_{c,1}/t=2.2$\cite{K13}.
Moreover, in Ref. \cite{afbf} it was shown that using {\em unrenormalized} 
energies of formation (i.e. $\kappa=0$) bond particle theory for the planar 
KLM did give the too large $J_c/t=2.20$ but reproduced the phase diagram of 
the model in the 
$(J/t,n_e)$-plane obtained by VMC\cite{WatanabeOgata} and
DMFT\cite{K13} quite well if $J$ was measured in units of $J_c$ - i.e. if the 
phase diagram was plotted in the $(J/J_c,n_e)$ plane - so that the error in 
$J_c/t$ cancelled to some degree. This is 
encouraging in that the phase diagram of the KLM in 2D is quite 
intricate, comprising the paramagnetic and two antiferromagnetic phases 
with different Fermi surface topology, with various 1$^{st}$ and 2$^{nd}$ order
transitions between them. Also, the band structure in the AF phase and its
change with $J/J_{c}$ as obtained by DCA\cite{MartinBerxAssaad} could be
reproduced in this way\cite{afbf1}. It should also be noted that in
the above bond particle calculations antiferromagnetic order appears
without an additional Heisenberg exchange between $f$ spins, that means it
comes about solely by the interaction mediated by the conduction electrons.

\end{document}